\documentclass{aa}  
\usepackage{graphicx}
\usepackage{natbib,txfonts}
\begin{document}
   \title{First VLTI infrared spectro-interferometry on \object{GCIRS~7}\thanks{Based on observations collected at the European Southern
     Observatory, Paranal, Chile (programs 076.B-0863, 077.D-0709).}}

   \subtitle{Characterizing the prime reference source for Galactic center observations at highest angular resolution}

   \author{J.-U.~Pott
          \inst{1}\fnmsep\inst{2}\fnmsep\inst{3}
          \and
          A.~Eckart\inst{3}
	  \and A.~Glindemann\inst{4}
          \and S.~Kraus\inst{5}
          \and R.~Sch{\"o}del\inst{6}
          \and A.~M.~Ghez\inst{2}
	  \and J.~Woillez\inst{1}
	  \and G.~Weigelt\inst{5}
          }

   \offprints{J.-U. Pott}

   \institute{
   W.M. Keck Observatory (WMKO), CARA, 65-1120 Mamalahoa Hwy., Kamuela, HI-96743, USA\\
              \email{jpott@keck.hawaii.edu}
         \and
University of California Los Angeles (UCLA), Department of Astronomy, 430 Portola Plaza, Los Angeles, CA 90095-1547, USA
	 \and
   I. Physikalisches Institut (PH1), University of Cologne, Z{\" u}lpicher Str. 77, D-50937 K\"oln, Germany
	 \and
   European Southern Observatory (ESO), 
              Karl-Schwarzschild Str. 2, D-85748 Garching bei M\"unchen, Germany
	 \and
   Max-Planck-Institut f{\" u}r Radioastronomie (MPIfR), Auf dem H{\" u}gel 69, D-53121 Bonn, Germany
         \and
   Instituto de Astrof{\' i}sica de Andaluc{\' i}a (IAA), Camino Bajo de Hu{\' e}tor, 50, 18008 Granada, Spain
             }

   \date{Received <date> / Accepted <date>}

 
  \abstract
   {The massive black hole Sgr A* at the very center of the Galaxy, and its immediate stellar and non-stellar environment, have been studied in the past decade with increasing intensity and wavelength coverage, revealing surprising results. This research requires the highest angular resolution available to avoid source confusion and to study the physical properties of the objects.}
   {GCIRS~7 is the dominating star of the central cluster in the NIR, so it has been used as wavefront and astrometric reference.
Our studies investigate, for the first time, its properties at 2 and 10~$\mu$m using the VLTI. We aim at analyzing the suitability of GCIRS~7 as an interferometric phase-reference for the upcoming generation of dual-field facilities at optical interferometers.}
   {VLTI-AMBER and MIDI instruments were used to spatially resolve GCIRS~7 and to measure the wavelength dependence of the visibility using the low spectral resolution mode ($\lambda / \Delta \lambda \approx 30$) and projected baseline lengths of about 50~m, resulting in an angular resolution of about 9~mas and 45~mas for the NIR and MIR, respectively.}
   {The first $K$-band fringe detection of a GC star suggests that GCIRS 7 could be marginally resolved at 2 micron, which would imply that the photosphere of the supergiant is enshrouded by a molecular and dusty envelope.
At 10~$\mu$m, GCIRS~7 is strongly resolved with a visibility of approximately 0.2. 
The MIR is dominated by moderately warm (200~K), extended dust, mostly distributed outside of a radius of about 120~AU (15~mas) around the star. 
A deep 9.8$\,\mu$m-silicate absorption in excess of the usual extinction law with respect to the NIR extinction has been found.}
   {Our VLTI observations show that interferometric NIR phase-referencing experiments with mas resolution using GCIRS~7 as phase reference appear to be feasible, but more such studies are required to definitely characterize the close environment around this star. The MIR data confirm recent findings of a relatively enhanced, interstellar 9.8$\,\mu$m-silicate absorption with respect to the NIR extinction towards another star in the central arc-seconds, suggesting an unusual dust composition in that region. 
We demonstrate that the resolution and sensitivity of modern large-aperture optical telescope arrays is required to resolve the innermost environment of stars at the Galactic center. }


   \maketitle
%

\section{Introduction}
Due to its proximity, the Galactic center (GC) offers the only opportunity to spatially resolve astrophysical phenomena in the immediate vicinity of a
massive black hole \citep[MBH;][]{1996Natur.383..415E,1998ApJ...509..678G}
and provides seminal knowledge for understanding spatially unresolved extra-galactic nuclei.
For the past decade, it has been possible to achieve diffraction-limited observations of the GC with 8-10m class telescopes, and therefore to probe the details of processes, such as star and dust formation at galacto-centric radii where the $\sim 4 \cdot 10^6\,M_\odot$ black hole generates a strong tidal field, at a unique angular resolution of about 8\,AU/mas \citep{2003ApJ...586L.127G,2005ApJ...628..246E}.
However, even at this resolution source confusion is a significant effect.
With sufficient intensity sensitivity, optical long baseline interferometry (OLBI) offers today, in the pre-ELT era, the capability to study this region at even higher angular resolution, increased by about an order of magnitude over single-telescope experiments. 
Large apertures as offered by the VLTI\footnote{{\bf VLT I}nterferometer: {\tt http://www.eso.org/projects/vlti/}, in this article we consider the array of 8m-UT's as VLTI, since currently the VLTI-AT's are not sensitive enough for GC observations.} and the Keck Interferometer (KI \footnote{{\tt http://planetquest.jpl.nasa.gov/Keck/keck\_index.cfm}}) have provided a breakthrough in sensitivity enabling first interferometric studies at 10~$\mu$m of the brightest ($N\,\gtrsim 1$~Jy) GC sources \citep{2005Msngr.119...43P}.

In the NIR the sensitivity constraints are even tighter due to significantly shorter atmospheric coherence times, and higher sensibility to instrumental vibrations along the optical path ($K_{\rm lim}\,\sim10$~mag). 
If such a bright and mostly unresolved source within the isoplanatic patch of the science target can be observed with the interferometer, the atmospheric piston-noise can be monitored enabling longer integration times and boosting the sensitivity, equivalent to natural-guide-star adaptive optics for single telescopes. 
Such phase-referencing facilities are in the making (VLTI: PRIMA, GRAVITY; KI: ASTRA), and have anticipated limiting $K$-magnitudes of about $15\,-19$. 
These facilities should allow studies of \object{Sgr A*}, which is the emission source associated with the MBH \citep[e.g.][]{2003Natur.425..934G}, and of orbits deeper in the gravitational potential of the MBH, which could test general relativity in the strong gravity regime \citep{2001A&A...374...95R, 2005ApJ...622..878W}.
It, therefore, appears to be timely to characterize the primary candidate of phase-referencing experiments at the GC, \object{GCIRS~7}~\footnote{In the following we will mostly use the abbreviation IRS~7.}.

This M1 supergiant \citep{1996AJ....112.1988B} is the brightest near-infrared star in the central parsec ($K\,\sim 7$~mag), and is located only 6~$\arcsec$ away from the MBH.
While photometric variability of IRS~7 implies a possible change of its diameter (and visibility) over time \citep{1996ApJ...470..864B,1999ApJ...523..248O}, the average $K$ magnitude suggests an average stellar photospheric diameter of about 1~mas at NIR wavelengths.
IRS~7 is therefore expected to be at the very limit of resolution for VLTI and KI, with maximum baseline lengths of 130~m and 85~m, respectively, and should be suitably compact to serve as an OLBI phase-reference source for the GC until regular fringe tracking on fainter stars becomes possible as proposed for the GRAVITY instrument \citep{2006SPIE.6268E..33G}. 

While single-dish observations of IRS~7 show it to be point-like at NIR wavelengths, similar observations in the MIR wavelengths regime reveal that a small fraction of the light is from an extended dust distribution. Specifically, \citet{2007A&A...462L...1S} detect in an 8.6~$\mu$m-image a tail-like structure directed away from the center of the galaxy \citep[see Fig.~1 in ][]{2007arXiv0711.0249P}.
This raises the concern that hotter dust may contribute substantially to the source size at 2~$\mu$m.
Furthermore, such features are interesting in their own right, as they suggest an ongoing interaction of the circumstellar dust with impinging external stellar winds from other stars, probably from the IRS~16 region of blue, hot, and windy stars \citep{1997A&A...325..700N}.

Probing the MIR source size as a function of wavelength over the $N$-band also probes the location of the dust giving rise to the high line-of-sight extinction in that band, dominated by the silicate absorption, as was done for \object{GCIRS~3} \citep{2007arXiv0711.0249P}. 
In that article, we presented first observational evidence, based on the exceptional angular resolution of an optical interferometer, for an unusual dust chemistry and silicate overabundance in the interstellar dust around GCIRS~3, located (projected) 200~mpc away of the MBH only, and the here presented MIDI experiment was designed to extend this evidence by observing another star in that region.

In this article, we report on our recent effort to confirm experimentally the compactness and symmetry of IRS 7 at $K$-band wavelengths  and at OLBI
baselines as a necessary preparation for future OLBI imaging and astrometric observations of the GC. 
In addition, first MIR-fringes on IRS~7, taken with the VLTI-MIDI instrument, are presented and discussed.
In the following section the observations, and data reduction process are described, including a discussion of the measurement uncertainties. Sect.~\ref{sec:3} discusses the merits of the presented data in the appropriate scientific context, separately for the $K$- and $N$-band. The main conclusions are summarized in Sect.~\ref{sec:4}.


\section{\label{sec:2}Observations and data reduction}
 
\begin{table}
\begin{minipage}[t]{\columnwidth}
\caption{Log of observations of the presented data.}
\label{tab:1}
\centering
\renewcommand{\footnoterule}{}  
\begin{tabular}{l c c c c c c}
\hline \hline
Target & UT  & PB & PA & Airm.\footnote{Airmass of the observation} & See.
\footnote{\label{foot:1}DIMM seeing towards zenith, and the coherence time $\tau_0$; the atmospheric data are taken from the ESO ambient conditions database: {\tt http://archive.eso.org/asm/ambient-server?site=paranal}}
    & $\tau_0$~\textsuperscript{\ref{foot:1}} \\    
    & & [m] & [degr] && [``] & [ms] \\
\hline
\multicolumn{2}{r}{AMBER-LR} & \multicolumn{5}{l}{band: 2.05-2.35~$\mu$m; $R=$30}\\
&&  \multicolumn{5}{l}{baseline: UT3-UT4; night: 14/03/2006} \\
\hline
\object{GCIRS7}~\footnote{\label{foot:6}We used \object{USNO-A2.0 0600-28579500} as an off-axis optical, natural AO guide star; B$\,\sim15.2$~mag, with 33~$''$ distance to IRS~7. }  & 08:27 & 51.8 & 94$^\circ$  &1.2 & 1.2& 6 \\
\object{HD153368}\footnote{Visibility-calibrator: $(1.01~\pm 0.01)$~mas~\textsuperscript{\ref{foot:5}}, K$_{2{\rm MASS}}$ = 3.4~mag \citep{2003tmc..book.....C}} & 
               09:08 & 59.7 & 101$^\circ$ &1.1& 0.9 & 9 \\
\hline
\multicolumn{2}{r}{MIDI-LR} & \multicolumn{5}{l}{band: 8.3-13.0~$\mu$m; $R=$30}\\
\multicolumn{2}{r}{mode:  {\sc High-sens}}&  \multicolumn{5}{l}{baseline: UT2-UT3; night: 13/06/2006} \\
\hline
\object{GCIRS~7}~\textsuperscript{\ref{foot:6}}& 03:37 & 46.3 & 29$^\circ$& 1.06 & 0.85 & 2.5 \\
\object{HD165135}\footnote{\label{foot:3}Visibility- and flux calibrator: $(3.4~\pm 0.2)$~mas~\textsuperscript{\ref{foot:5}}, K0III, $F\, (12~\mu m)\,=\,15.5~Jy$ \citep[color-corrected IRAS flux;][]{1988iras....1.....B,1994yCat.2125....0J}}&04:16 & 46.5 & 32$^\circ$& 1.04 & 0.9  & 2.5 \\
\hline
\end{tabular}
\end{minipage}
\end{table}

   \begin{figure}
   \centering
   \includegraphics[width=\columnwidth]{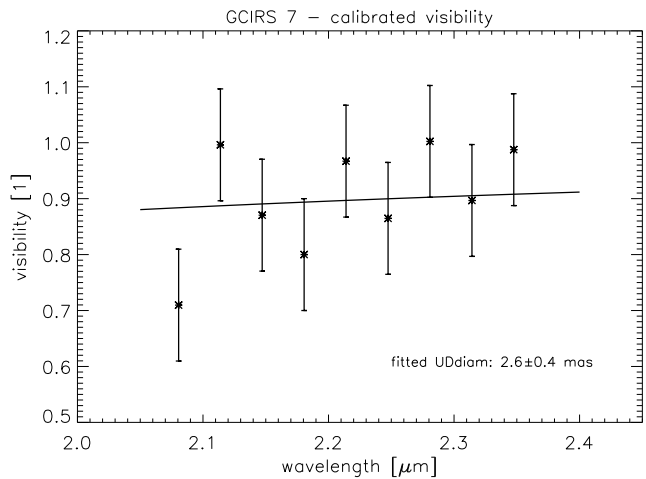}
      \caption{Calibrated $K$-band visibility of IRS~7. A uniform disc model is over-plotted (solid line), the fitted diameter and the statistic uncertainty is given.
              }
         \label{fig:1}
   \end{figure}

   \begin{figure}
   \centering
   \includegraphics[width=\columnwidth]{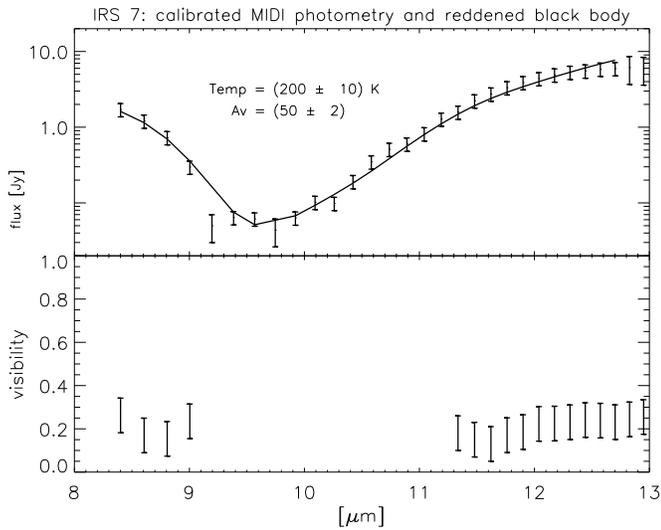}
      \caption{Calibrated total-flux spectrum and spectro-visibilities of IRS~7. The plotted flux errors show an absolute calibration uncertainty of about 20~\%. In the upper panel, the best-fit, reddened black-body is overplotted as solid line, the two fit-parameters are indicated with their 1~$\sigma$ uncertainties. 
         }
         \label{fig:2}
   \end{figure}

The first successful near infrared fringe detection on IRS~7 was achieved with the VLTI first generation science beam combiner AMBER \citep{2007A&A...464....1P} in March 2006 on the short UT3-UT4 baseline. 
In June 2006, the high-angular-resolution dataset could be extended by a MIDI UT2-UT3 observation. 
Each observation was followed by the measurement of a calibrator star of known diameter with negligible uncertainty with respect to the precision of our observations, given in Table~\ref{tab:1}, taken from ESO's observations preparation pages \footnote{\label{foot:5}See the {\sc CalVin} tool on {\tt http://www.eso.org/observing/etc}}, and measured with the identical instrumental setup of the IRS~7 observations, to calibrate for systematic instrumental and atmospheric visibility corruptions. 
A complete log of all relevant VLTI observations is given in Table~\ref{tab:1}. 
\subsection {\label{sec:21}AMBER - $K$-band spectro-interferometry}
Since IRS~7 is close to the sensitivity limit of AMBER, we chose the low spectral resolution mode ($R\,\sim 30$) and a detector integration time (DIT) of 0.1~s for all observations. Being longer than the 25~ms standard, this DIT increases the flux per frame, and is still short enough with respect to the atmospheric coherence time (Table~\ref{tab:1}) to detect the fringe pattern, however at decreased precision due to the increased fringe blur due to atmospheric turbulence over one DIT.
The AMBER data were reduced with the {\sc Amdlib} package originally provided by the instrument consortium\footnote{{\tt http://www.mariotti.fr/data\_processing\_amber.htm}}. Details of the reduction process applied to similar data were recently discussed \citep[see][and references therein]{2007arXiv0711.4988K}. 
Deviating from the aforementioned article, we relaxed the frame-selection criteria by accepting flux ratios between the telescopes up to 10 instead of an often chosen ratio of about 4 to increase the number of acceptable frames to a statistically relevant number, on cost of the average precision of the individual fringe measurement. 

A total of 62 frames, out of all 2400 frames taken, targeting IRS~7, passed the selection criteria.
This exceptionally low observing efficiency of a success rate of about 3\% derives in part from an imperfect delay-line model, the interferometric equivalent of a pointing model, at the time of the observation which prohibited repeated fringe measurements at longer baselines than UT3-UT4 at all.
The now implemented, significantly improved model minimizes the number of fringe losses due to earth rotation and target motion over the night sky, and will increase the success rate of future experiments.
Furthermore, vibrations of the UT's reduce the fringe contrast \citep{2006Msngr.126...37B}. 
The presented observations suffer from these conditions in particular due to the short coherence length of the low-spectral-resolution mode, and due to a low SNR due to the intrinsically low source flux and the off-axis AO-guiding.

The appropriate fringe SNR  \citep{2007A&A...464...29T} was calculated for each accepted frame. 
To minimize the influence of possible systematics related to low SNR, we used only the frames from the best 30~percentile for the further analysis in this article, where {\it best} means highest SNR.
The visibility of these best frames were averaged without applying a weight.  
Using less frames would lead to too small frame numbers possibly biasing the final average.

The resulting average visibilities of our AMBER data are plotted with error-bars of 0.1 in  Fig.~\ref{fig:1}. This error resembles the variance in our data, but might underestimate the accuracy due to the relatively small number of accepted frames and systematic errors in the interferometric transfer function.
The  standard deviation of the final visibility data points around the best-fit uniform-disc model, over-plotted in the same figure, gives similar 10\%.
This precision is lower than visibility uncertainties of about 3~\% obtained with AMBER in other experiments \citep{2007A&A...464....1P}. 
A lower precision is expected here due to our relaxed frame selection criteria, the low source flux, and the relatively long DIT.
Furthermore, the data-point at 2.08~$\mu$m showing the lowest visibility of the ensemble might, in addition to the aforementioned sources of error, suffer a signal corruption from an imperfect wavelength-calibration and telluric lines around this wavelength.  Its inclusion does not significantly change the fit results presented in Sect.~\ref{sec:33}.

\subsection{MIDI - $N$-band spectro-interferometry}
Both calibrator and science measurement consist of a standard on-source exposure time of 2.4~min, subdivided into 8000 individual frame exposures of 18~ms, nearly all of which entered the final visibility calculation in contrast to the low frame-acceptance rate in the NIR. 
This different observing efficiency originates in the longer observing wavelength minimizing the impact of linear delay-errors along the optical path (e.g. evoked by vibrating optics) on the phase noise, and in shorter individual exposure times despite of a longer atmospheric coherence time in the MIR.
In contrast to turbulence-induced coherence-time constraints in the NIR, the short frame exposure times at 10~$\mu$m are set by the high thermal background noise level.

The spectrally dispersed MIDI data ($R\,\sim$ 30) were reduced with the publicly available MIA+EWS software package~\footnote{\label{foot:4}Used version from Dec-21-2007, available at: {\tt http://www.strw.leidenuniv.nl/$\sim$nevec/MIDI/index.html}}. We applied the error analysis developed in \citet{2007arXiv0711.0249P}, where we published similar data. 
The combination of low source flux due to high interstellar extinction and low atmospheric transmission around 10~$\mu$m leads to the lack of reliable visibility and flux data-points in the center of the $N$-band (Fig.~\ref{fig:2}). 
Additional sources of uncertainties in both the calibrated flux spectrum and dispersed visibility include the off-axis AO guiding (see Table~\ref{tab:1}), a slightly varying performance of one of the two beams during the observations, and the time difference between the correlated and total flux measurement of the used {\sc High-sens} mode, which was 9 min and 6 min for IRS~7 and the calibrator, respectively. 
Due to the lack of additional calibrator measurements during that particular night, the spectral shape of the flux density is probably better constrained than the absolute flux calibration (Fig.~\ref{fig:2}). 

The measured total-flux MIR spectrum was flux-calibrated with the spectrum of the visibility calibrator specified in Table~\ref{tab:1}, and is plotted in the upper panel in Fig.~\ref{fig:2}.
 A reddened black-body was fitted to the flux data. 
The 1-$\sigma$ uncertainties in the model parameters from a least-squares-fit are given in the figure.
The reduced $\chi ^2$ of 0.6 of the fit as well as coinciding integrated flux of both the model and the flux data demonstrate the goodness of the fit.
 We excluded the data below 50~mJy, and the ones beyond 12.8~$\mu$m, to avoid a corruption of the fit by these data-points.
The error of the fluxes below the 50~mJy limit is more than the 20\% flux error for data points used in the fit. 
The flux data-points at the longest wavelengths probably are not reliable and too low, because at those wavelengths the photometry data show a distorted PSF, and significantly different fluxes coming from each of the telescopes. 
Note that such a photometry bias does not necessarily affect the {\it normalized} visibility shown in the lower panel of the figure, as long as the conditions are stable during both the visibility and the photometry measurement.
 
\section{\label{sec:3}Discussion}

Since at the different wavelengths of our two VLTI datasets different temperatures, radii and presumably different types of emitting matter dominate the observed flux, we split the following discussion according to these wavelength bands.

\subsection{\label{sec:33}NIR} 

Our data suggest that IRS~7 could be resolved (i.e. has a visibility of less than unity) in the $K$-band at the short VLTI baseline used.
We fitted a wavelength-independent diameter of a uniform-disc model to the NIR-data. 
This model, adequate for the given accuracy and size of the data-set, minimizes the $\chi ^2$-metric ($\chi^2_{\rm red} \approx 0.9$) at $(2.6\,\pm\,0.4)$~mas, which is a remarkable result, because such a best-fit diameter is larger than expected.
However, the total error including systematics like the variation of the interferometric transfer function and the small number of interferograms can increase the uncertainty on the derived diameter up to the order of $\pm$1.5~mas (AMBER team, private communication). 
Repeated measurements at a similar setup, seeing, and instrument performance are required to further investigate in detail the systematics of the here used AMBER setup.

 To calculate the diameter of the 
photosphere of an IRS~7-like supergiant of early MI spectral type at the distance to the GC ($\sim$ 7.6 kpc), we assume $T_{\rm eff}\approx\,3400$~K, and a bolometric correction $BC_{\rm K}\,=2.7$ \citep{1996AJ....112.1988B,2003ApJ...597..323B}.
Recent PSF-fitted VLT photometry and extinction estimations are $K\,=\,7.4$ and $A_{\rm K}\,=\,3.7$ for a 2005 dataset of ours. 
This is about a magnitude fainter than the Blum values based on 1997 data. 
Since the typical precison for these estimates is about 0.1 mag for photometry and extinction, this discrepancy points to a significant photometric variability, and follows the trend reported by \citet{1999ApJ...523..248O}. 
Therefore we use our recent photometry  measurements here, being relatively close in time to our AMBER measurement.
 We derive a stellar radius of about (1000$\,\pm\,150)$~$R_\odot$ or an angular diameter of (1.1$\,\pm\,0.2$)~mas at the GC distance. 
Recent interferometric studies of $\alpha$ Ori, the nearby red supergiant \object{Betelgeuse} of spectral type M1-2Ia-Ibe similar to IRS~7, showed a larger measured radius at NIR (and MIR) wavelengths than the expected stellar photosphere due to a surrounding molecular gas shell of significantly lower temperature (about 2000~K) and a radius of about 1.4 times the radius of the stellar photosphere \citep{2004A&A...421.1149O,2007A&A...474..599P}.

The  increased $\sim$1.6$\,\pm\,0.2$~mas diameter of such an $\alpha$~Ori-like, warm molecular gas shell  around the photosphere of IRS~7 would still be significantly smaller than the uniform-disc diameter derived here from the AMBER data, but the respective theoretical visibility of ($0.96\,\pm\,0.01$) lies within the total error of our observation.
 Without a more precise dataset of repeated AMBER measurements allowing for cross-calibration to minimize this total error of about 1.5~mas we cannot determine the significance of contributions of other (and larger) shells than the photosphere to the NIR flux of IRS~7.
However, astrophysics also can be responsible for intrinsically lower than expected visibilities.
The visibility drops, if in addition to the flux of a  putative circumstellar molecular shell the radiation of more extended warm dust has been resolved out by the interferometer \citep[e.g. observed in the supergiant \object{IRC~+10420},][]{1999A&A...348..805B}. 
Indeed, a significant flux contribution  of cooler, extended dust is seen in the MIDI data discussed below.
Furthermore, a change in optical depth due to molecular band-heads between 2.2 and 2.4~$\mu$m can contribute to an apparent size increase at these wavelengths.

In terms of phase-referencing several aspects have to be considered. 
As long as the visibility is high enough the SNR of the correlated flux is high and a stable phase-reference is provided.
But here we show data on the shortest VLTI baseline only.
The uniform-disc model (2.4~mas diameter) predicts a visibility of 0.45 at the longest UT-baseline (130~m). 
That means, only half of the correlated flux will be seen at such long baselines reducing the SNR, if the model is applicable.
Furthermore, any resolvable asymmetry in the brightness distribution of IRS~7 would lead to non-zero closure-phases and a complex visibility function.
This hampers the usage IRS~7 as phase-reference unless the position-angle dependence of this phase is known.
Furthermore if indeed warm molecular gas and dust shells make up the NIR-visibility of IRS~7, temporal variations of optical depth and radial size in such shell can result in significant visibility changes and would have to be known if IRS~7 shall be used as visibility calibrator in a dual-field experiment. 
 More observational data similar to the one presented here is needed to answer these questions in detail. But since IRS~7 is by far the brightest star in the NIR, located within the central 5~$\arcsec$, and according to our results compact enough to give a strong interferometric signal on a 50~m baseline, we can confirm it to be the prime candidate for future dual-star phase-referencing experiments in the GC.

\subsection{MIR}

As suggested by previous VLT/VISIR imaging and mentioned in the introduction, IRS~7 {\it is} surrounded by significant amounts of warm dust, radiating in the MIR. 
It is however surprising that our MIDI experiment revealed average $N$-band visibilities as low as 0.2. 
Lacking detailed a priori knowledge about the dust morphology of IRS~7, observed at a nominal interferometric angular resolution of about 45~mas, such visibility amplitudes can be interpreted in two related ways focusing on the flux measurement and on the spatial scales probed.

(1) A flat visibility spectrum as in Fig.~\ref{fig:2} is created by an unresolved, (circum-)stellar point source radiating 20\% of the total flux plus a more extended and smooth brightness distribution the flux of which is completely resolved out by tbe interferometer, and appears only in the total flux spectrum (upper panel in Fig.~\ref{fig:2}). 
Assuming a Gaussian intensity profile for the extended flux with the constraint of contributing less than 10\% of its flux to the visibility measurement leads to a FWHM of 36~mas at 10~$\mu$m.
This interpretation suggests that, apart from the central emission, about 80\% of the total flux is (smoothly) radiated from (projected) distances greater than 150~AU from the star.

(2) Another common interpretation aims on estimating an average angular scale-length of the entire flux distribution by fitting a Gaussian profile to the observed visibility measurements. 
The measured flat visibility spectrum results in a wavelength-dependent FWHM: 24, 30, and 39~mas at 8, 10, and 13~$\mu$m, respectively. 
An increasing diameter towards longer wavelengths is expected for optically thin, thermal sources showing cooler temperatures, which dominate the flux at longer wavelengths, further out. 

Summarizing, the MIR-brightness distribution of IRS~7 is clearly dominated by dust emission extended on 30~mas scales and greater. 
Accordingly, the fitted black-body temperature of only 200~K (Fig.~\ref{fig:2}) appears to fit the complete $N$-band emission. This (for close stellar environments) relatively low dust temperature is another confirmation of observing dust relatively far away from the central heat source. 

A possible explanation of the origin of this extended dust emission (and of the low visibilities) derives from radio wavelengths. 
More than 15~years ago \citet{1991ApJ...371L..59Y} spatially resolved in 2~cm VLA maps of 0.4~$\arcsec$ angular resolution a bow-shock-like free-free radio emission region around IRS~7. 
The curved emission morphology shields IRS~7, which lies $\sim$150~mas North of the apex of the shock front according to the model of \citet{1992ApJ...385L..41Y}, from the heating and ionizing galactic winds emanating from the very center of the Galaxy.  
It is conceivable that (a part of) the extended MIR emission detected by MIDI is generated in this externally heated environment.
The radiative dominance of this emission over the warmer dust closer to the star might point to the outer dust either being optically thick, or overwhelming the inner, slightly hotter dust closer to IRS~7 due to an energetic domination of the external radiation field over heating by IRS~7 itself. 

In line with this interpretation is that the position angle of the projected baseline during the time of observation is with $PA\,\approx\,30^\circ$ (East-of-North) relatively close to the roughly North-South elongation of the radio bow-shock.
This would  explain the low amount of correlated MIR-flux. 
A better $uv$-coverage is needed to prove this scenario. 
Estimating the putative apex and radiative properties of a MIR shock front around IRS~7 pointing to the GC will allow us to estimate physical properties complementing the radio data, and to investigate the interstellar conditions only a few arc-seconds away from the MBH.

The calibrated total-flux spectrum from the MIDI photometry of IRS~7 shows a broad interstellar 9.8$\,\mu$m-silicate absorption feature.
The optical depth $\tau_{\rm 9.8~\mu m}\,\approx\,7$ in this feature is twice as high as expected for the usually adopted interstellar MIR extinction towards the central parsec, as derived from earlier observations of the GC at much lower angular resolution.
This $\tau_{\rm 9.8~\mu m}$ resembles our recent findings towards the nearby IRS~3 \citep{2007arXiv0711.0249P}.
 This strong $\tau_{\rm 9.8~\mu m}$-excess cannot solely be attributed to circumstellar dust enshrouding only these two stars for various reasons.
The spectral shape of the 9.8~$\mu$m-silicate feature fits the typical shape of {\em interstellar} silicate dust absorption.
The NIR-continuum extinction towards these two stars \citep{2007A&A...469..125S}  does not show a similar excess, and circumstellar silicate dust normally shows the silicate feature in emission. 
This  paradox of normal extinction at NIR and relatively enhanced $\tau_{\rm 9.8~\mu m}$ suggests a relative silicate overabundance over carbonaceous dust grains in the diffuse interstellar environment of the entire central parsec, or at least around IRS~3 and IRS~7.
Further support comes from the MIDI spectro-visibilities of IRS~7.
The essentially flat curve means that the dust resolved by the interferometer, and located {\em close} to the star, does {\em not} account for a major fraction of the excess of interstellar silicate absorption, since otherwise the change in optical depth along the silicate feature would be visible as a spectral feature in the visibility spectrum.
Here again, 'close to the star' means closer than about 30~mas. 

The flat visibility over the $N$-band thus suggests that the dust responsible for the $\tau_{\rm 9.8~\mu m}$-excess is located further away from the star, diffusely distributed within the central parsec, and suggests an unusual silicate overabundance in the GC-dust chemistry.
That the, usually linear, relation between the optical depth in the silicate feature and the extinction at other wavelengths, known from various studies and locations throughout the galaxy, can break down in regions with particular properties has recently been confirmed for dense interstellar clouds \citep{2007ApJ...666L..73C}. 
In the Galactic center, recent observations report the possibility of an unusual stellar initial mass function \citep[e.g.][]{2007MNRAS.374L..29K,2007ApJ...669.1024M}, which could affect the GC dust chemistry.
 Also \citet{2007ApJ...669.1011C} derived from chemical abundance measurements of 10 GC stellar atmospheres within the central 30~pc enhanced [O/Fe] and [Ca/Fe] ratios, and concluded an overabundance of $\alpha$-elements in the GC, including Si. This could explain an interstellar $\tau_{\rm 9.8~\mu m}$-excess with respect to the carbon-dominated NIR-extinction, as suggested by our MIDI data.

\section{\label{sec:4}Conclusion}
We present the first successful NIR fringe measurements with an optical long baseline interferometer of a star in the Galactic center at a nominal angular resolution of 9~mas.
The analysis of the 50~m-baseline $K$-band visibilities of the red supergiant \object{IRS~7} leads to a uniform-disc diameter of 2.6~mas which suggests that all the VLTI-UT baselines up to a length of 130~m resolve the starlight.
This size  is larger than the expected diameter of the photosphere and suggests the detection of hot circumstellar dust and molecular shells. 
But more NIR-data at high SNR are needed to confirm this result, and to estimate an accurate diameter of IRS~7.  The strong interferometric signal observed shows the feasibility of future interferometric phase-referencing experiment involving IRS~7.

First MIR interferometric data taken on IRS~7 with the VLTI-MIDI instrument proved that most of the flux at 10~$\mu$m stems from moderately warm, extended dust, possibly due to shock-interaction with external stellar winds dominating the MIR flux. 
A de-reddening of the MIDI spectra revealed high silicate absorption in excess of standard GC values, similar to the nearby IRS~3. A local overabundance of interstellar, amorphous silicate throughout the central arc-secs is suggested.

\begin{acknowledgements}
The professional support of the ESO Paranal VLTI team, guaranteeing
efficient technically advanced observations, is acknowledged.
This research has made use of the SIMBAD database,
operated at CDS, Strasbourg, France. Part of this work was supported
by the German {\em Deutsche Forschungsgemeinschaft} (DFG via SFB 494) and by an ESO
studentship (JUP). Further JUP gratefully acknowledges the hospitality and support of the IR-IF group at MPIfR (G.~Weigelt) and the GC group at UCLA (A.~Ghez).
\end{acknowledgements}

\bibliography{gcirs7}

\begin{thebibliography}{33}
\expandafter\ifx\csname natexlab\endcsname\relax\def\natexlab#1{#1}\fi

\bibitem[{{Beichman} {et~al.}(1988){Beichman}, {Neugebauer}, {Habing}, {Clegg},
  \& {Chester}}]{1988iras....1.....B}
{Beichman}, C.~A., {Neugebauer}, G., {Habing}, H.~J., {Clegg}, P.~E., \&
  {Chester}, T.~J., eds. 1988, {Infrared astronomical satellite (IRAS) catalogs
  and atlases. Volume 1: Explanatory supplement}, Vol.~1

\bibitem[{{Bl{\"o}cker} {et~al.}(1999){Bl{\"o}cker}, {Balega}, {Hofmann},
  {Lichtenth{\"a}ler}, {Osterbart}, \& {Weigelt}}]{1999A&A...348..805B}
{Bl{\"o}cker}, T., {Balega}, Y., {Hofmann}, K.-H., {et~al.} 1999, \aap, 348,
  805

\bibitem[{{Blum} {et~al.}(2003){Blum}, {Ram{\'{\i}}rez}, {Sellgren}, \&
  {Olsen}}]{2003ApJ...597..323B}
{Blum}, R.~D., {Ram{\'{\i}}rez}, S.~V., {Sellgren}, K., \& {Olsen}, K. 2003,
  \apj, 597, 323

\bibitem[{{Blum} {et~al.}(1996{\natexlab{a}}){Blum}, {Sellgren}, \&
  {Depoy}}]{1996ApJ...470..864B}
{Blum}, R.~D., {Sellgren}, K., \& {Depoy}, D.~L. 1996{\natexlab{a}}, \apj, 470,
  864

\bibitem[{{Blum} {et~al.}(1996{\natexlab{b}}){Blum}, {Sellgren}, \&
  {Depoy}}]{1996AJ....112.1988B}
{Blum}, R.~D., {Sellgren}, K., \& {Depoy}, D.~L. 1996{\natexlab{b}}, \aj, 112,
  1988

\bibitem[{{Bonnet} {et~al.}(2006){Bonnet}, {Bauvir}, {Wallander}, {Cantzler},
  {Carstens}, {Caruso}, {di Lieto}, {Guisard}, {Haguenauer}, {Housen},
  {Mornhinweg}, {Nicoud}, {Ramirez}, {Sahlmann}, {Vasisht}, {Wehner}, \&
  {Zagal}}]{2006Msngr.126...37B}
{Bonnet}, H., {Bauvir}, B., {Wallander}, A., {et~al.} 2006, The Messenger, 126,
  37

\bibitem[{{Chiar} {et~al.}(2007){Chiar}, {Ennico}, {Pendleton}, {Boogert},
  {Greene}, {Knez}, {Lada}, {Roellig}, {Tielens}, {Werner}, \&
  {Whittet}}]{2007ApJ...666L..73C}
{Chiar}, J.~E., {Ennico}, K., {Pendleton}, Y.~J., {et~al.} 2007, \apjl, 666,
  L73

\bibitem[{{Cunha} {et~al.}(2007){Cunha}, {Sellgren}, {Smith}, {Ramirez},
  {Blum}, \& {Terndrup}}]{2007ApJ...669.1011C}
{Cunha}, K., {Sellgren}, K., {Smith}, V.~V., {et~al.} 2007, \apj, 669, 1011

\bibitem[{{Cutri} {et~al.}(2003){Cutri}, {Skrutskie}, {van Dyk}, {Beichman},
  {Carpenter}, {Chester}, {Cambresy}, {Evans}, {Fowler}, {Gizis}, {Howard},
  {Huchra}, {Jarrett}, {Kopan}, {Kirkpatrick}, {Light}, {Marsh}, {McCallon},
  {Schneider}, {Stiening}, {Sykes}, {Weinberg}, {Wheaton}, {Wheelock}, \&
  {Zacarias}}]{2003tmc..book.....C}
{Cutri}, R.~M., {Skrutskie}, M.~F., {van Dyk}, S., {et~al.} 2003, {2MASS All
  Sky Catalog of point sources.}

\bibitem[{{Eckart} \& {Genzel}(1996)}]{1996Natur.383..415E}
{Eckart}, A. \& {Genzel}, R. 1996, \nat, 383, 415

\bibitem[{{Eisenhauer} {et~al.}(2005){Eisenhauer}, {Genzel}, {Alexander},
  {Abuter}, {Paumard}, {Ott}, {Gilbert}, {Gillessen}, {Horrobin}, {Trippe},
  {Bonnet}, {Dumas}, {Hubin}, {Kaufer}, {Kissler-Patig}, {Monnet},
  {Str{\"o}bele}, {Szeifert}, {Eckart}, {Sch{\"o}del}, \&
  {Zucker}}]{2005ApJ...628..246E}
{Eisenhauer}, F., {Genzel}, R., {Alexander}, T., {et~al.} 2005, \apj, 628, 246

\bibitem[{{Genzel} {et~al.}(2003){Genzel}, {Sch{\"o}del}, {Ott}, {Eckart},
  {Alexander}, {Lacombe}, {Rouan}, \& {Aschenbach}}]{2003Natur.425..934G}
{Genzel}, R., {Sch{\"o}del}, R., {Ott}, T., {et~al.} 2003, \nat, 425, 934

\bibitem[{{Ghez} {et~al.}(2003){Ghez}, {Duch{\^e}ne}, {Matthews}, {Hornstein},
  {Tanner}, {Larkin}, {Morris}, {Becklin}, {Salim}, {Kremenek}, {Thompson},
  {Soifer}, {Neugebauer}, \& {McLean}}]{2003ApJ...586L.127G}
{Ghez}, A.~M., {Duch{\^e}ne}, G., {Matthews}, K., {et~al.} 2003, \apjl, 586,
  L127

\bibitem[{{Ghez} {et~al.}(1998){Ghez}, {Klein}, {Morris}, \&
  {Becklin}}]{1998ApJ...509..678G}
{Ghez}, A.~M., {Klein}, B.~L., {Morris}, M., \& {Becklin}, E.~E. 1998, \apj,
  509, 678

\bibitem[{{Gillessen} {et~al.}(2006){Gillessen}, {Perrin}, {Brandner},
  {Straubmeier}, {Eisenhauer}, {Rabien}, {Eckart}, {Lena}, {Genzel}, {Paumard},
  \& {Hippler}}]{2006SPIE.6268E..33G}
{Gillessen}, S., {Perrin}, G., {Brandner}, W., {et~al.} 2006, SPIE, 6268, 33

\bibitem[{{IRAS PSC Version 2.0}(1994)}]{1994yCat.2125....0J}
{IRAS PSC Version 2.0}. 1994, VizieR Online Data Catalog, 2125, 0

\bibitem[{{Klessen} {et~al.}(2007){Klessen}, {Spaans}, \&
  {Jappsen}}]{2007MNRAS.374L..29K}
{Klessen}, R.~S., {Spaans}, M., \& {Jappsen}, A.-K. 2007, \mnras, 374, L29

\bibitem[{{Kraus} {et~al.}(2008){Kraus}, {Preibisch}, \&
  {Ohnaka}}]{2007arXiv0711.4988K}
{Kraus}, S., {Preibisch}, T., \& {Ohnaka}, K. 2008, ApJ in press, ArXiv
  e-prints 0711.4988

\bibitem[{{Maness} {et~al.}(2007){Maness}, {Martins}, {Trippe}, {Genzel},
  {Graham}, {Sheehy}, {Salaris}, {Gillessen}, {Alexander}, {Paumard}, {Ott},
  {Abuter}, \& {Eisenhauer}}]{2007ApJ...669.1024M}
{Maness}, H., {Martins}, F., {Trippe}, S., {et~al.} 2007, \apj, 669, 1024

\bibitem[{{Najarro} {et~al.}(1997){Najarro}, {Krabbe}, {Genzel}, {Lutz},
  {Kudritzki}, \& {Hillier}}]{1997A&A...325..700N}
{Najarro}, F., {Krabbe}, A., {Genzel}, R., {et~al.} 1997, \aap, 325, 700

\bibitem[{{Ohnaka}(2004)}]{2004A&A...421.1149O}
{Ohnaka}, K. 2004, \aap, 421, 1149

\bibitem[{{Ott} {et~al.}(1999){Ott}, {Eckart}, \&
  {Genzel}}]{1999ApJ...523..248O}
{Ott}, T., {Eckart}, A., \& {Genzel}, R. 1999, \apj, 523, 248

\bibitem[{{Perrin} {et~al.}(2007){Perrin}, {Verhoelst}, {Ridgway}, {Cami},
  {Nguyen}, {Chesneau}, {Lopez}, {Leinert}, \&
  {Richichi}}]{2007A&A...474..599P}
{Perrin}, G., {Verhoelst}, T., {Ridgway}, S.~T., {et~al.} 2007, \aap, 474, 599

\bibitem[{{Petrov} {et~al.}(2007){Petrov}, {Malbet}, {Weigelt}, {Antonelli},
  {Beckmann}, {Bresson}, {Chelli}, {Dugu{\'e}}, {Duvert}, {Gennari},
  {Gl{\"u}ck}, {Kern}, {Lagarde}, {Le Coarer}, {Lisi}, {Millour}, {Perraut},
  {Puget}, {Rantakyr{\"o}}, {Robbe-Dubois}, {Roussel}, {Salinari}, {Tatulli},
  {Zins}, {Accardo}, {Acke}, {Agabi}, {Altariba}, {Arezki}, {Aristidi},
  {Baffa}, {Behrend}, {Bl{\"o}cker}, {Bonhomme}, {Busoni}, {Cassaing},
  {Clausse}, {Colin}, {Connot}, {Delboulb{\'e}}, {Domiciano de Souza},
  {Driebe}, {Feautrier}, {Ferruzzi}, {Forveille}, {Fossat}, {Foy},
  {Fraix-Burnet}, {Gallardo}, {Giani}, {Gil}, {Glentzlin}, {Heiden},
  {Heininger}, {Hernandez Utrera}, {Hofmann}, {Kamm}, {Kiekebusch}, {Kraus},
  {Le Contel}, {Le Contel}, {Lesourd}, {Lopez}, {Lopez}, {Magnard}, {Marconi},
  {Mars}, {Martinot-Lagarde}, {Mathias}, {M{\`e}ge}, {Monin}, {Mouillet},
  {Mourard}, {Nussbaum}, {Ohnaka}, {Pacheco}, {Perrier}, {Rabbia}, {Rebattu},
  {Reynaud}, {Richichi}, {Robini}, {Sacchettini}, {Schertl}, {Sch{\"o}ller},
  {Solscheid}, {Spang}, {Stee}, {Stefanini}, {Tallon}, {Tallon-Bosc}, {Tasso},
  {Testi}, {Vakili}, {von der L{\"u}he}, {Valtier}, {Vannier}, \&
  {Ventura}}]{2007A&A...464....1P}
{Petrov}, R.~G., {Malbet}, F., {Weigelt}, G., {et~al.} 2007, \aap, 464, 1

\bibitem[{{Pott} {et~al.}(2008){Pott}, {Eckart}, {Glindemann}, {Sch{\"o}del},
  {Viehmann}, \& {Robberto}}]{2007arXiv0711.0249P}
{Pott}, J.-U., {Eckart}, A., {Glindemann}, A., {et~al.} 2008, \aap, 480, 115

\bibitem[{{Pott} {et~al.}(2005){Pott}, {Eckart}, {Glindemann}, {Viehmann},
  {Schodel}, {Straubmeier}, {Leinert}, {Feldt}, {Genzel}, \&
  {Robberto}}]{2005Msngr.119...43P}
{Pott}, J.-U., {Eckart}, A., {Glindemann}, A., {et~al.} 2005, The Messenger,
  119, 43

\bibitem[{{Rubilar} \& {Eckart}(2001)}]{2001A&A...374...95R}
{Rubilar}, G.~F. \& {Eckart}, A. 2001, \aap, 374, 95

\bibitem[{{Sch{\"o}del} {et~al.}(2007{\natexlab{a}}){Sch{\"o}del}, {Eckart},
  {Alexander}, {Merritt}, {Genzel}, {Sternberg}, {Meyer}, {Kul}, {Moultaka},
  {Ott}, \& {Straubmeier}}]{2007A&A...469..125S}
{Sch{\"o}del}, R., {Eckart}, A., {Alexander}, T., {et~al.} 2007{\natexlab{a}},
  \aap, 469, 125

\bibitem[{{Sch{\"o}del} {et~al.}(2007{\natexlab{b}}){Sch{\"o}del}, {Eckart},
  {Mu{\v z}i{\'c}}, {Meyer}, {Viehmann}, \& {Bower}}]{2007A&A...462L...1S}
{Sch{\"o}del}, R., {Eckart}, A., {Mu{\v z}i{\'c}}, K., {et~al.}
  2007{\natexlab{b}}, \aap, 462, L1

\bibitem[{{Tatulli} {et~al.}(2007){Tatulli}, {Millour}, {Chelli}, {Duvert},
  {Acke}, {Hernandez Utrera}, {Hofmann}, {Kraus}, {Malbet}, {M{\`e}ge},
  {Petrov}, {Vannier}, {Zins}, {Antonelli}, {Beckmann}, {Bresson}, {Dugu{\'e}},
  {Gennari}, {Gl{\"u}ck}, {Kern}, {Lagarde}, {Le Coarer}, {Lisi}, {Perraut},
  {Puget}, {Rantakyr{\"o}}, {Robbe-Dubois}, {Roussel}, {Weigelt}, {Accardo},
  {Agabi}, {Altariba}, {Arezki}, {Aristidi}, {Baffa}, {Behrend}, {Bl{\"o}cker},
  {Bonhomme}, {Busoni}, {Cassaing}, {Clausse}, {Colin}, {Connot},
  {Delboulb{\'e}}, {Domiciano de Souza}, {Driebe}, {Feautrier}, {Ferruzzi},
  {Forveille}, {Fossat}, {Foy}, {Fraix-Burnet}, {Gallardo}, {Giani}, {Gil},
  {Glentzlin}, {Heiden}, {Heininger}, {Kamm}, {Kiekebusch}, {Le Contel}, {Le
  Contel}, {Lesourd}, {Lopez}, {Lopez}, {Magnard}, {Marconi}, {Mars},
  {Martinot-Lagarde}, {Mathias}, {Monin}, {Mouillet}, {Mourard}, {Nussbaum},
  {Ohnaka}, {Pacheco}, {Perrier}, {Rabbia}, {Rebattu}, {Reynaud}, {Richichi},
  {Robini}, {Sacchettini}, {Schertl}, {Sch{\"o}ller}, {Solscheid}, {Spang},
  {Stee}, {Stefanini}, {Tallon}, {Tallon-Bosc}, {Tasso}, {Testi}, {Vakili},
  {von der L{\"u}he}, {Valtier}, \& {Ventura}}]{2007A&A...464...29T}
{Tatulli}, E., {Millour}, F., {Chelli}, A., {et~al.} 2007, \aap, 464, 29

\bibitem[{{Weinberg} {et~al.}(2005){Weinberg}, {Milosavljevi{\'c}}, \&
  {Ghez}}]{2005ApJ...622..878W}
{Weinberg}, N.~N., {Milosavljevi{\'c}}, M., \& {Ghez}, A.~M. 2005, \apj, 622,
  878

\bibitem[{{Yusef-Zadeh} \& {Melia}(1992)}]{1992ApJ...385L..41Y}
{Yusef-Zadeh}, F. \& {Melia}, F. 1992, \apjl, 385, L41

\bibitem[{{Yusef-Zadeh} \& {Morris}(1991)}]{1991ApJ...371L..59Y}
{Yusef-Zadeh}, F. \& {Morris}, M. 1991, \apjl, 371, L59

\end{thebibliography}
\bibliographystyle{aa}

\end{document}